\documentclass[leqno,openbib]{article}
\usepackage{float}
\usepackage{lineno}
\usepackage{url}
\usepackage[utf8]{inputenc}
\usepackage{booktabs}
\usepackage{indentfirst}
\usepackage{amsfonts}
\usepackage{ntheorem}
\usepackage{amsmath}
\usepackage[singlespacing]{setspace}
\usepackage{sectsty}
\usepackage[norule,bottom]{footmisc}
\usepackage[justification=centering,textfont={sc},labelfont={rm}]{caption}
\usepackage{varioref}
\usepackage{natbib}
\setcitestyle{authoryear,open={(},close={)}}
\usepackage[labelfont=bf]{caption}
\usepackage [autostyle, english = american]{csquotes}
\MakeOuterQuote{"}
\usepackage{graphicx}

\theoremindent\parindent
\makeatletter    
\renewtheoremstyle{plain}{\item[\hskip\labelsep\hskip-\parindent \theorem@headerfont ##1\ ##2\theorem@separator]}{\item[\hskip\labelsep\hskip-\parindent \theorem@headerfont ##1\ ##2\ (##3)\theorem@separator]}
\makeatother
\theoremheaderfont{\scshape}
\theorembodyfont{\rmfamily}
\theoremseparator{.}

\sectionfont{\normalfont\scshape\centering}

\subsectionfont{\itshape}
\makeatletter
\def\@biblabel#1{\hspace*{-\labelsep}}

\makeatother
\makeatletter
\renewcommand\@makefnmark{\mbox{\textsuperscript{\normalfont\@thefnmark}}}
\renewcommand\@makefntext[1]{\indent\makebox[2.5em][r]{\@thefnmark.\,}#1}
\if@titlepage
  \renewenvironment{abstract}{      \titlepage
      \null\vfil
      \@beginparpenalty\@lowpenalty
      \begin{center}        \@endparpenalty\@M
      \end{center}}     {\par\vfil\null\endtitlepage}
\else
  \renewenvironment{abstract}{      \if@twocolumn
      \else
        \small
        \begin{center}        \end{center}      \fi}
      {\if@twocolumn\else\endquotation\fi}
\fi
\renewcommand\thetable{\@Roman\c@table}

\renewcommand\thefigure{\@Roman\c@figure}

\makeatother
\labelformat{equation}{(#1)}
\begin{document}

\title{FAIR ODDS FOR NOISY PROBABILITIES}
\author{
\begin{tabular}{c}
\textsc{Ulrik W. Nash} \\ 
\textsc{Syddansk Universitet} \\ 
\textsc{5230 Odense M, Denmark.} \\
\textsc{Email: uwn@sam.sdu.dk} \\
\end{tabular}
\textsc{}}
\maketitle

\begin{abstract}
We suggest that one individual holds multiple degrees of belief about an outcome, given the evidence. We then investigate the implications of such noisy probabilities for a buyer and a seller of binary options and find the odds agreed upon to ensure zero-expectation betting, differ from those consistent with the relative frequency of outcomes. More precisely, the buyer and the seller agree to odds that are higher (lower) than the reciprocal of their averaged unbiased probabilities when this average indicates the outcome is more (less) likely to occur than chance. The favorite-longshot bias thereby emerges to establish the foundation of an equitable market. As corollaries, our work suggests the old-established way of revealing someone's degree of belief through wagers may be more problematic than previously thought, and implies that betting markets cannot generally promise to support rational decisions.\\
\textbf{JEL-codes:} D81, D83, D84, D87. \\
\textbf{PsycINFO classification:} 2340. \\
\end{abstract}

\thispagestyle{empty}\setcounter{page}{0}%
\newpage 

\section{Introduction}

Gambling, broadly defined, appears to characterize our lives \citep{FrankPRamsey1926, Reichenbach1938a, Brunswik1943a, Savage1954}: given bounded rationality and time constraints, we choose among alternatives based on limited degrees of belief and face the prospect of adverse outcomes as a consequence. Under these conditions, we strive to reduce our surprise by gathering evidence for and against alternatives while trying to manage the trade-off between decision speed and accuracy. Economists have long recognized the importance of these twin phenomena \citep{Stigler1961, Arrow1973} and watched institutions emerge to support the production and distribution of knowledge \citep{Machlup1962}. However, with modern technology, scholars are now taking deeper vantage points and have discovered rivalling hierarchies of neurons in the brain, which support alternative hypotheses, gather evidence towards these, and compete to decide the choices we make \citep{Shadlen1996, Heekeren2008}.

Neurons positioned upstream in these hierarchies modulate their firing rate in proportion to the strength of evidence for their supported alternative, while neurons downstream modulate their average firing rate as if accumulating the evidence detected \citep{Newsome1989, Britten1993}. Moreover, the response times of subjects are generally longer for weaker evidence and shorter for stronger evidence, suggesting that part of the solution developed by the brain works to balance the speed and accuracy of decisions by requiring neurons to reach thresholds for activity, indicative of confidence \citep{Roitman2002, Palmer2005}.

What makes the stated discovery particularly fascinating is what it suggests. It suggests the brain has evolved to perform something like sequential analysis, (SA) invented independently by Turing to break Enigma during World War II, and by Wald to determine if munitions were of a satisfactory quality to ship by the US Army during the same conflict \citep{Gold2002a}. This method involves a direct mathematical relation between the weight of evidence for a hypothesis and the probability of the hypothesis being correct \citep{Good1985a}, which suggests that science is starting to unravel how our brains update the beliefs \citep{FrankPRamsey1926, DeFinetti1937, Edwards1963} on which we wager in the face of uncertainty \citep{Kiani2009, Kiani2014}. 

In SA, new evidence and updated probabilities follow each other deterministically through Bayes' factor in the string of samples. Moreover, different sequences through the whole evidence ultimately lead to same posterior probability distribution. But in the neurophysiological counterpart, the coupling between evidence and probability appears less constrained. The stylized phenomenon of across-trial variability in choices by an individual presented with the same evidence \citep{Faisal2008, Pisauro2017,Ratcliff2016a} suggests that noise between the presented evidence and its internal representation by neurons makes the posterior probability distribution different from the one produced by the precision of mathematics. 

While SA computes one probability for each possible outcome, given the sampled evidence, it would appear the neurophysiological counterpart produces not one probability for each possible outcome, but an entire distribution of probabilities, given the same evidence. For example, while SA might ultimately suggest a 60 percent probability of Boston Red Socks winning, multiple judgments by the same individual are possible on the same evidence. Consequently, were this individual to bet on the match, their appraisal of fair odds might correspondingly vary, despite appearing definite at the moment of communication. On the other hand, one might also argue that even SA can have this effect, since there are many sequences through the whole evidence for and against alternative hypotheses, and because the process of SA may be stopped as time runs out. Whatever the case might be, however, our basic suggestion remains: subjective probabilities are not only dependent on the evidence, but also on the method by which we process that evidence, and that may add degrees of randomness to degrees of beliefs.

The purpose of this paper is to examine the consequence of noisy probabilities for odds agreed between a buyer and a seller of binary options, that is to say, between two gamblers defined in the way most commonly understood. Our predictions suggest that noisy, but unbiased probabilities, compel these gamblers to drive a wedge between the odds that are consistent with the relative frequencies of possible outcomes, and the odds that are consistent with zero-expectation bets. In particular, the buyer and the seller must agree to odds that are \textbf{higher} than the reciprocal of their averaged unbiased beliefs when this average indicates the outcome is \textbf{more} plausible than chance, and agree to odds that are \textbf{lower} than the reciprocal of their averaged unbiased beliefs when this average indicates the outcome is \textbf{less} plausible than chance. The gamblers must do this in order to cancel an unfairness towards the seller in odds greater than two, and cancel an unfairness towards the buyer in odds smaller than two, which exists when probabilities are noisy, and gamblers place equal weight on their respective beliefs.

Our predictions have, therefore, several important implications for economics. First, they suggest the definition of fair odds must account for the level of noise in probabilities held by each side of the market. Second, our predictions explain the favorite-longshot bias without involving preferences for risk among individual gamblers \citep{Friedman1952}, without involving misrepresentations of probabilities by individuals \citep{Kahneman1979b}, and without involving asymmetric information between them \citep{Shin1992}. Instead, our predictions suggest the favorite-longshot bias is an equilibrium outcome that establishes the foundation of a fair market. Third, our work suggests noisy probabilities make the old-established way of revealing degrees of belief through wagering more difficult than previously thought \citep{FrankPRamsey1926,DeFinetti1937,Carnap1962,Edwards1963}, which, finally, casts doubt on the promise of betting markets to provide unbiased indicators of future events and support rational economic behavior \citep{Arrow}.

\section{The Basic Game}
Sunday has come around again, and you are at the match once more with your friends. You have read the latest news on both teams, and from what you can gather, your probability of the home team  winning is fifty-fifty. You have no reason to think any of your friends are less informed than you, and still, you cannot wait to continue the betting scheme you have been following for some time with success. This week, Bob is your unwitting target.

"Hey Bob," you say as you nudge him in the ribs with your elbow. "I have an excellent proposition for you. Do you see this small box? It contains my probability of our team winning today. What I want you to do is state your probability to me, open the box, and look at what I have written. Then we calculate the simple average of our two numbers and wager at the odds it implies. Our probabilities will also decide which side of the market each of us takes. If your probability is greater than mine, you back our team and take the role of buyer, while I become your bookmaker and take the role of the seller; otherwise, you take that role. I propose we stake one dollar. What do you say?"

Bob nods. "Sure, why not?"

"Excellent, but there is a final condition to my proposal. Once we see the odds and know which side of the market we have, we are free to abandon."

To Bob, your proposition sounds unproblematic, except for the last condition, which made him think. Given the match program sticking out of your top pocket, he knows you have read the latest news, but so has he, which should rule out the problem of asymmetric information. Moreover, the lecture in statistics covering the classic thought experiment by \citet{DeFinetti1937} on how to reveal someone's degree of belief wasn't one that Bob missed, and to him, your arrangement seems close to that setting. Except there was no sealed probability in De Finetti's story, or perhaps there was, but it remained sealed and discounted completely. Moreover, taking the average of beliefs across the market is intuitively fair to Bob, and agrees with what he knows about the wisdom of crowds \citep{Galton1907}, which has been said to characterize betting markets \citep{Arrow}. Since you and he will be forming a small market, this argument comforts Bob. As for the abandon clause, he cannot see its effect and even finds it mildly insulting; Bob might be an economics student, but he can certainly afford to lose a few multiples of a dollar!

"Right, you're on! Bob exclaims. My probability is 0.3".

"Splendid, Bob," you reply and request that Bob now opens the small box. 

"As you see, my probability is 0.5, and our consensus is therefore 0.4, which means the odds we agree are one over 0.4, or 2.5. Also, since my probability is higher than yours, I back our team, and you offer the odds as my bookmaker. So, to be clear, if our team loses, I will give you \$1  dollar, and otherwise, you give me \$1.50. Does that sound fair to you, Bob?" 

"Yes, it does. And I am not bailing out!"  

"Me neither, Bob, me neither."

\section{Modelling the basic game}

Let us model the basic game by first considering two independent mechanisms that produce probabilities of mutually exclusive binary outcomes. In other words, let us consider two independent mechanisms that estimate the relative frequency by which independent binary outcomes would occur if events could be repeated endlessly at constant initial conditions. Both mechanisms function by sequentially detecting and accumulating evidence \citep{Good1985a, Gold2002a, Gold2007a} for and against the competing hypotheses, $h{_{1}}$ and  $h{_{0}}$, before converting the evidence to pairs of numbers on the unit interval. Formally, we assume each mechanism uses the weight of evidence to generate probabilities $P_{h{_1}}$ and $P_{h{_0}}$ of the respective hypotheses being correct \citep{Gold2002a}

\begin{equation}
\label{eq:1}
P_{h{_1}} =\frac{1}{10^{-WOE_{h_1}}+1}, ~~~ P_{h{_0}} =1 - P_{h{_1}}, 
\end{equation}
where the weight of evidence gathered $WOE_{h_1}$ is expressed in bans following the convention used by Turing \citep{Good1985a}. Note from probability equation \ref{eq:1} that since evidence \textbf{for} one hypothesis is evidence \textbf{against} its alternative, the sum of any probability pair never fails to equal one, and as such, no opponent can ever establish a Dutch Book against either mechanism.

\section{The Baseline Solution}

Let us now assume two players of the basic game use the stated mechanisms to form degrees of belief about the binary outcomes. We also assume the players have utility functions that are linear with money, have no market power, and wager only when their subjective expected margin ($\pi_B{_s}$ in the role of the buyer, and $\pi_S{_s}$ in the role of the seller) is non-negative. 

In the baseline solution, which we derive for regular comparison, probabilities are noiseless and odds $\frac{1}{P_C}$ are found by equating the subjective expected margin on each side of the market

\begin{equation}
\label{eq:2}
\pi_B{_s}=\left(\frac{1}{P_C}-1\right) P_B-\left(1-P_B\right) \\
\end{equation}
and
\begin{equation}
\label{eq:3}
\pi_S{_s}=\left(1-P_S\right)-\left(\frac{1}{P_C}-1\right) P_S,
\end{equation}
\noindent to yield the consensus
\begin{equation}
\label{eq:4}
P_C=\frac{1}{2} \left(P_B+P_S\right).
\end{equation}
We notice solution \ref{eq:4} corresponds to the pricing rule in the basic game, but there are a few other things to notice too. First, solution \ref{eq:4} provides market observers with an unbiased estimate of the relative frequency $P_T$ observable if events could be repeated endlessly at constant initial conditions. Second, the solution is more reliable than $P_B$ and $P_S$ individually and thereby harnesses the wisdom of crowds, precisely as Bob noted, and finally, since solution \ref{eq:4} places equal weight on the beliefs, it seems fair.

Nevertheless, as we demonstrate below, when two players with noisy probabilities play the basic game repeatedly, they will sooner or later find it unfair; although both players would break-even across all wagers eventually, their margin in the role of buyer would be positive while their margin in the role of seller would be negative. Accordingly, we notice the abandon clause in the basic game serves to offer protection against negative expectation bets for the seller.

\section*{Introducing Noisy Probabilities}
Let us be specific about the nature of stochasticity in the evidence accumulation process, and then examine why the basic game becomes unfair. We assume probabilities are produced uniformly across the appropriate range $0 \leq P_T \leq 1$, thereby making things as simple as possible while maintaining the assumption that $P_B$ and $P_S$ are unbiased estimates of $P_T$. Accordingly, the continuous uniform distribution

\begin{equation}
\label{eq:5}
P_B, P_S \sim \mathcal{U}(L,H),
\end{equation}

\noindent describes $P_B$ and $P_S$, where $L = P_T - E$, $H = P_T + E$, $E = \epsilon \cdot min(1 - P_T, P_T)$ and $0\leq \epsilon \leq 1$ (Figure 1). As such, $\epsilon$ is a tunable parameter that can be used to set noise to the maximum level governed by the range between $L$ and $H$ (i.e., $\epsilon = 1$), eliminate noise completely (i.e., $\epsilon = 0$), adjust noise to somewhere in between, while ensuring probabilities stay within their required limits and remain an unbiased estimate of $P_T$.

\begin{figure}
    \centering
    \textbf{The Assumed Character of Noise:}\par\medskip
    \includegraphics[scale=1]{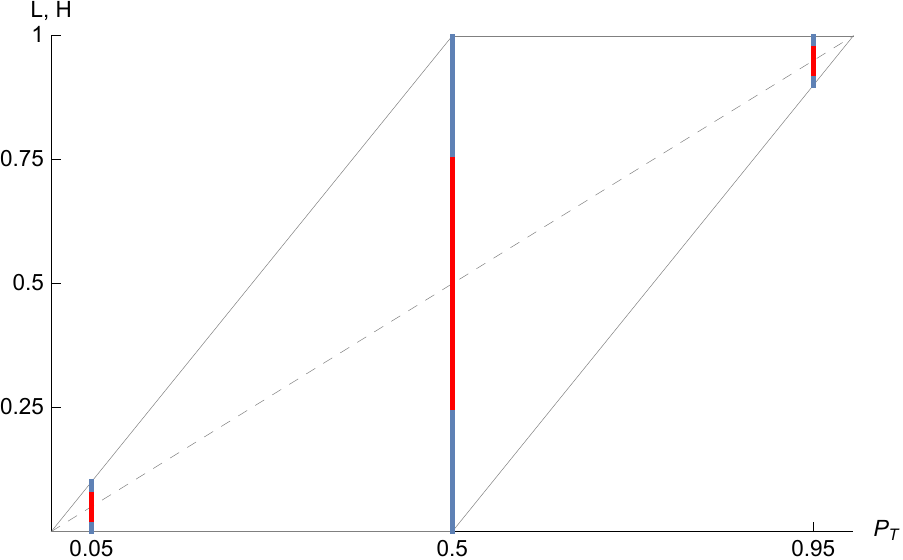}
    \caption{\normalfont For each magnitude of relative frequency $P_T$ (i.e., the objective probability), the buyer's probabilities $P_B$ and the seller's probabilities $P_S$ (i.e., their degrees of belief about $P_T$) are produced uniformly and without bias. How noisy the probabilities are is determined by what proportion $\epsilon$ of the uniform range around $P_T$ from $L$ to $H$ the probabilities are drawn from. The thick red lines indicate $\epsilon= 0.50$, while the thick blue lines show $\epsilon= 1$,  at $P_T = 0.05$, $P_T = 0.50$, and $P_T = 0.95$ respectively.}
\end{figure}

Given these assumptions, we can now state how the gathered weight of evidence must couple to the whole evidence $WOE_T$, without needing to model the accumulation process explicitly. Rearranging equation \ref{eq:1} to obtain the equation for weight of evidence as a function of probability, we first obtain

\begin{equation}
\label{eq:1b}
WOE_{h_1} = -\frac{\log \left(\frac{1-P_{h_1}}{P_{h_1}}\right)}{\log (10)}, ~P_{h_1} \sim \mathcal{U}(L,H)
\end{equation}

Next, denoting
\begin{equation}
\label{eq:6}
a=\ln \left(\frac{1}{H}-1\right)+P_{h{_1}} \ln (10), 
\end{equation}
and
\begin{equation}
\label{eq:7}
b=\ln \left(\frac{1}{L}-1\right)+P_{h{_1}}\ln (10), 
\end{equation}
we find the gamblers gather $WOE_{h_1}$ described (Figure 2, left) by the cumulative distribution function \\ 

\begin{equation}
\label{eq:8}
F(P_{h{_1}}) = {\begin {cases} 0 & : L>0\land a>0 \\ \frac{10^{P_{h{_1}}}}{\left(10^{P_{h{_1}}}+1\right) (H-L)} & : L=0\land a\leq 0 \\ \frac{H}{H-L} & : L=0\land a>0 \\ \frac{1 -2 L + \tanh \left(\frac{1}{2} P_{h{_1}} \ln (10)\right)}{2 (H-L)} & : L>0\land a\leq 0\land b>0 \\ 1 & : \text{Otherwise,} \end {cases}}\\
\end{equation}
whose corresponding probability density function,
\begin{equation}
\label{eq:9}
f(P_{h{_1}}) = \frac{10^{P_{h{_1}}} \ln (10)}{\left(10^{P_{h{_1}}}+1\right)^2 (H-L)},
\end{equation}
has a median corresponding to probability equation \ref{eq:1} (Figure 2, right). Accordingly, we surmise the evidence accumulation mechanisms produce unbiased estimates of $P_T$ because they tend to gather the whole evidence, albeit unreliably.

\begin{figure}
    \centering
    \textbf{The Implied Character of Evidence Accumulation:}\par\medskip
    \includegraphics[scale=1]{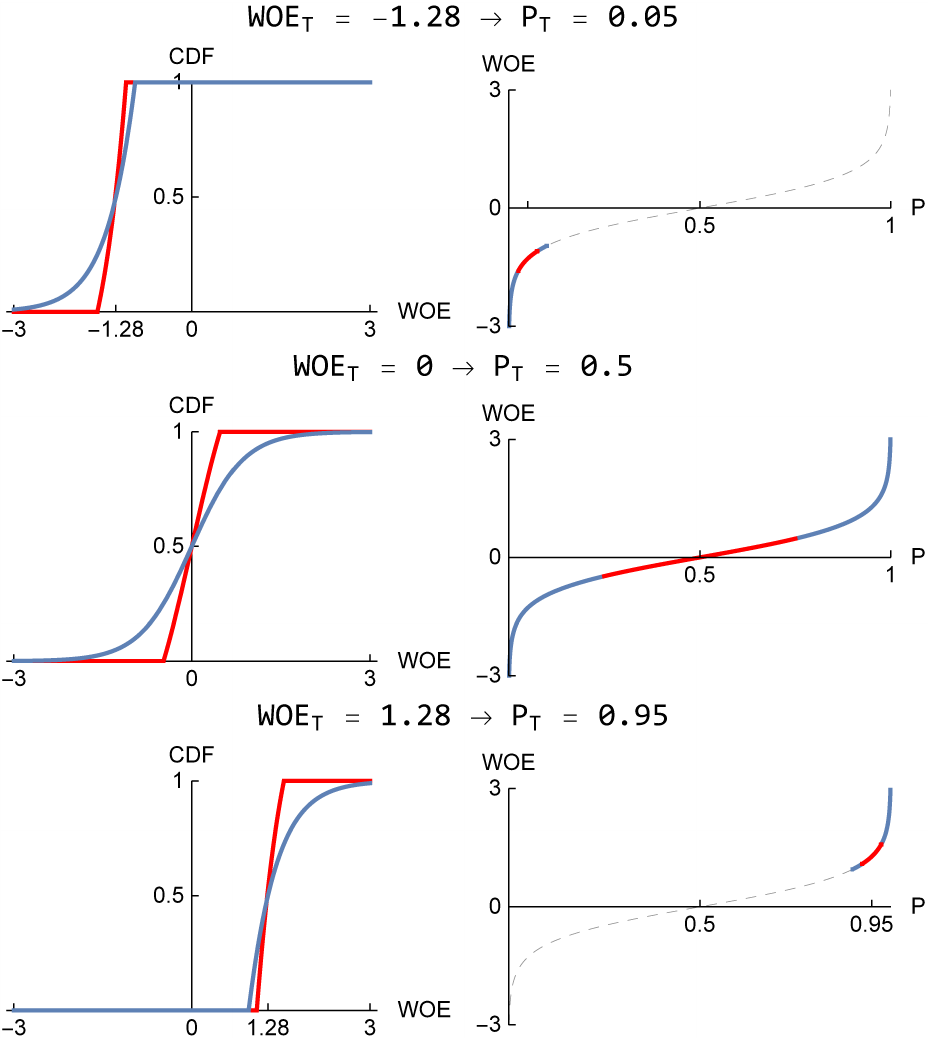}
    \caption{\normalfont  Left: The given stimulus condition is characterized by the whole weight of evidence $WOE_T$. However, we deduce the weight of evidence gathered from must be distributed according to the CDF of equation \ref{eq:8} for our assumptions about $P_B$ and $P_S$ to hold. We show the cases $(WOE_T = -1.28, P_T = 0.05)$, $(WOE_T = 0, P_T = 0.50)$ and $(WOE_T = 1.28, P_T = 0.95)$ in continuation of Figure 1. Right: The relation between noisy evidence accumulation and noisy probabilities is shown explicitly using equation \ref{eq:1}. As in Figure I, the red lines concern $\epsilon = 0.5$ while blue lines show $\epsilon = 1$  }
\end{figure}

\section{The Unfairness of the Basic Game}
The basic game is fair when players gather the whole evidence faultlessly, but not when they gather evidence unreliably. Given the zero-sum nature of the basic game, we can show this simply by focusing on the role of the seller. Keeping things general by assigning the weight $w_S$ to the seller's probability and $w_B = 1 - w_S$ to the buyer's probability, we begin by stating the seller's expected margin as

\begin{equation}
\label{eq:10}
\pi _{S_o}=\left(1-P_T\right)-P_T \left(\frac{1}{P_B \left(1-w_S\right)+P_S w_S}-1\right)
\end{equation}

\noindent and then condition it on the role allocation rule $P_S < P_B$, before computing the mean of \ref{eq:10} across this subset of transactions. Denoting

\begin{eqnarray}
\label{eq:11}
c=-P_T(1-\epsilon )\\
\label{eq:12}
d=-P_T(\epsilon +1)\\
\label{eq:13}
g=2 P_T \epsilon \\
\label{eq:14}
h=w_1(g-2 \epsilon )+\epsilon-c,
\end{eqnarray}

\noindent and applying the stated condition, we find the mean of the seller's expected margin splits at the point of chance along the spectrum of $P_T$, which we show using subscripts $1$ and $2$ in 

\begin{equation}
\label{eq:15}
\bar{\pi }_{S_{o_1}} = 1 + \frac{\frac{c \ln \left(\frac{c}{d+g w_1}\right)}{1-w_1}+\frac{d \left(\ln (d)-\ln \left(d+g w_1\right)\right)}{w_1}}{g \epsilon}
\end{equation}

\noindent for $0\leq P_T < \frac{1}{2}$, and 

\begin{equation}
\label{eq:16}
\bar{\pi }_{S_{o_2}} = 1 + \frac{\frac{(d+\epsilon) (\ln (-(d+\epsilon))-\ln (h))}{1-w_1}-\frac{(\epsilon-c) (\ln (\epsilon-c)-\ln (h))}{w_1}}{g \epsilon \left(-\frac{4 \epsilon}{g}+\frac{1}{P_T^2}+1\right)}
\end{equation}

\noindent for $\frac{1}{2} \leq P_T \leq 1$. 

Equation $\ref{eq:15}$ reveals that when players set odds using the arithmetic average of their probabilities (solution \ref{eq:4}), and then gamble indiscriminately at these odds, they create a situation where the seller eventually realizes a negative margin for all values of $P_T$ except $P_T = 1$ ($\lim_{P_T\to 1} \, \bar{\pi }_{S_{o_B}} = 0$), and for all values of noise except $\epsilon = 0$ ($\lim_{\epsilon \to 0} \, \bar{\pi }_{S_{o_1}} = 0$, $\lim_{\epsilon \to 0} \, \bar{\pi }_{S_{o_2}} = 0$). 

At the special point $w_1 = \frac{1}{2}$, $\epsilon = 1$, and $P_T$ approaching $\frac{1}{2}$, the value of $\bar{\pi }_{S_{o_2}}$ is $\frac{\ln \left(1-w_1\right)}{w_1}+1$ or $-0.39$. But more generally, $\bar{\pi }_{S_{o_1}}$ is negative for $0 < \epsilon \leq 1$ when $w_1 = \frac{1}{2}$, and indeed any $0 < w_1 < 1$. These predictions can be appreciated visually from the horizontal line's upward movement in the top and middle rows of Figure 3 as noise decreases. As for the range $\frac{1}{2} \leq P_T \leq 1$ captured by equation \ref{eq:16}, whatever value $\bar{\pi }_{S_{o_2}}$ takes at $P_T = \frac{1}{2}$, the same margin will eventually be earned across the entire range $0 \leq P_T \leq \frac{1}{2}$, since $\frac{\partial \bar{\pi }_{S_{o_1}} }{\partial P_T} = 0$ in that interval. Moreover, with any $0 \leq \epsilon \leq 1$ and $0 \leq w_1 \leq \frac{1}{2}$, the value of $\bar{\pi }_{S_{o_2}}$ approaches but never enters the positive region, as $P_T$ rises above $P_T = \frac{1}{2}$ towards $1$. However, with enough weight placed on the buyer's belief, $\frac{1}{2} < w_1 \leq 1$, the seller secures a positive margin over the range $\frac{1}{2} < P_T < 1$. 

\begin{figure}
    \centering
    \textbf{The Unfairness of the Basic Game:}\par\medskip
    \includegraphics[scale=1]{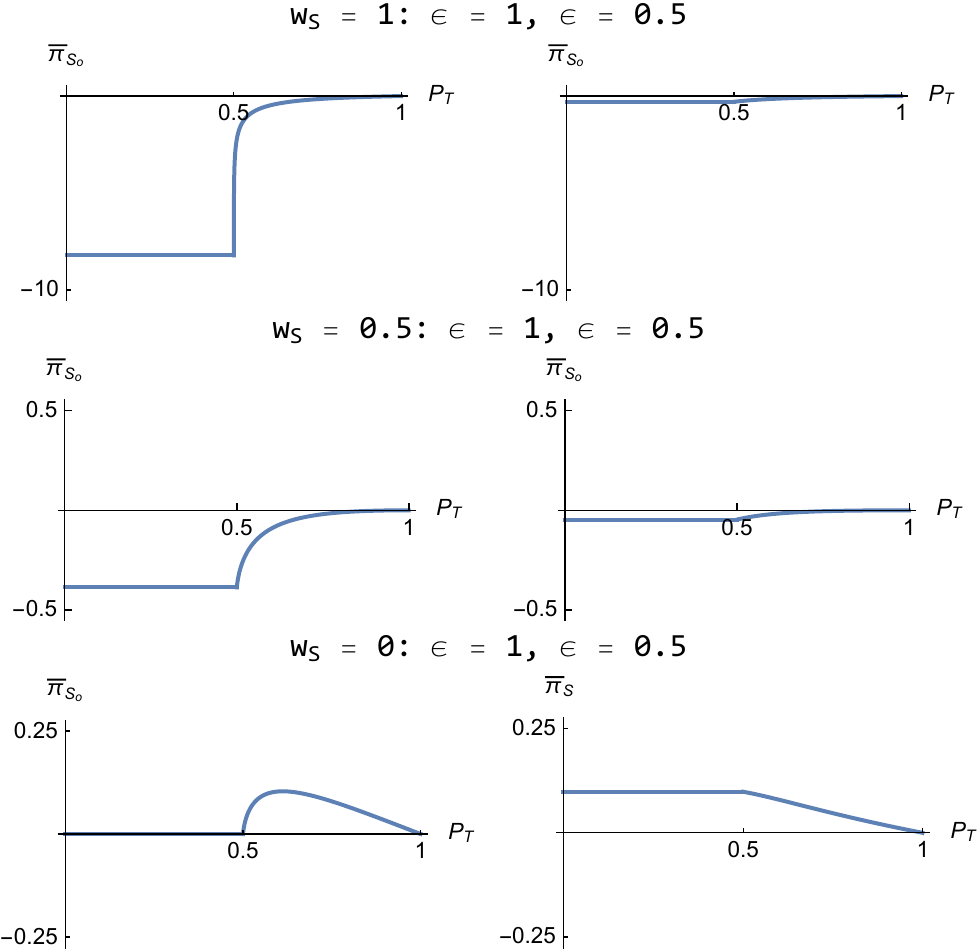}
    \caption{\normalfont When the buyer and the seller agree to odds by placing equal weight on their respective probabilities, as described by \ref{eq:4} in the basic game, then an economic asymmetry between them arises. For $0 < \epsilon \leq 1$ and $0 < P_T < 1$, the mean expected margin for the seller from gambling indiscriminately at the agreed odds is negative, while the corresponding value for the buyer is positive due to the zero-sum nature of the basic game. In short, in the basic game, the buyer and the seller are predicted to have symmetric economic conditions only when probabilities are noiseless.}
\end{figure}

In short, when players have noisy probabilities, set odds using solution \ref{eq:4}, and gamble indiscriminately at these odds, they should trigger the abandon-clause when given the role of seller, which means the basic game ends when both players reach this conclusion. 

\subsection*{The Origin of Unfairness}
At the origin of unfairness towards the role of seller is an asymmetry between the cost for the seller of $P_C$ underestimating $P_T$, compared to the benefit of $P_C$ overestimating $P_T$ by the same magnitude. To see this, let us define $\iota$ as $|P_T - P_C|$, such that $P_C = P_T + \iota$ is beneficial for the seller, whereas $P_C = P_T - \iota$ is costly. The seller's margin in the beneficial case is consequently given by $\left(1-P_T\right)-P_T \left(\frac{1}{P_T + \iota}-1\right)$, whereas the seller's margin in the costly instance is given by $\left(1-P_T\right)-P_T \left(\frac{1}{P_T - \iota}-1\right)$. Subtracting the latter from the former now reveals the asymmetry

\begin{equation}
\label{eq:17}
\Delta =-\frac{2 \iota  P_T}{\iota ^2-P_T^2},
\end{equation}
which never enters the positive region, since $\iota$ must be smaller or equal to $P_T$ in order for $P_C = P_T - \iota$ to remain non-negative.

\section*{Making the Basic Game Fair}
In order for gambling to continue, the buyer must compensate the seller for $\Delta$, and we find it reasonable to think players of the basic game will seek alternatives to solution \ref{eq:4} as they begin to sense $\Delta$ through their financial performance. In the following, we examine two candidate amendments to the basic game and predict that only the second is feasible.

\subsection*{Candidate A: Unequally Weighted Beliefs}
Since the equal weight on beliefs required by solution \ref{eq:4} becomes unfair when probabilities are noisy, it seems natural to presume that players of the basic game will begin to converge on odds that place most weight on the belief held by the buyer. In other words, for each relative frequency the buyer and seller will set $w_1$ such that odds $\frac{1}{P_C}$ makes neither part worse off when accepting these odds indiscriminately. Unfortunately, no symbolic solution exists for this attempt, but as illustrated by the left column of Figure 4, the numeric value of $w_1$ that yields $\bar{\pi }_{S_{o_1}} = 0$ and $\bar{\pi }_{S_{o_2}} = 0$ for different settings of $P_T$ and $\epsilon$ is available. We denote this weight by $w_1^*$.

When the buyer and the seller have increasingly noisy degrees of belief about $P_T$, then $w_1^*$ decreases for all $0 \leq P_T < 1$, thereby creating consensus odds that are smaller than $\frac{1}{P_T}$ on that range. Moreover, the difference between $\frac{1}{P_C}$ and $\frac{1}{P_T}$ is greater for $P_T < \frac{1}{2}$ (Figure 4, right bottom). Unfortunately, while players would discover they can achieve break-even \textbf{if} they remain in agreement to gamble indiscriminately at odds determined by $w_1^*$, this agreement cannot be sustained since indiscriminate gambling at odds determined by $w_1^*$ is suboptimal behaviour.

\begin{figure}
    \centering
    \textbf{Candidate Solution A: Unequal Weighting of Beliefs:}\par\medskip
    \includegraphics[scale=0.75]{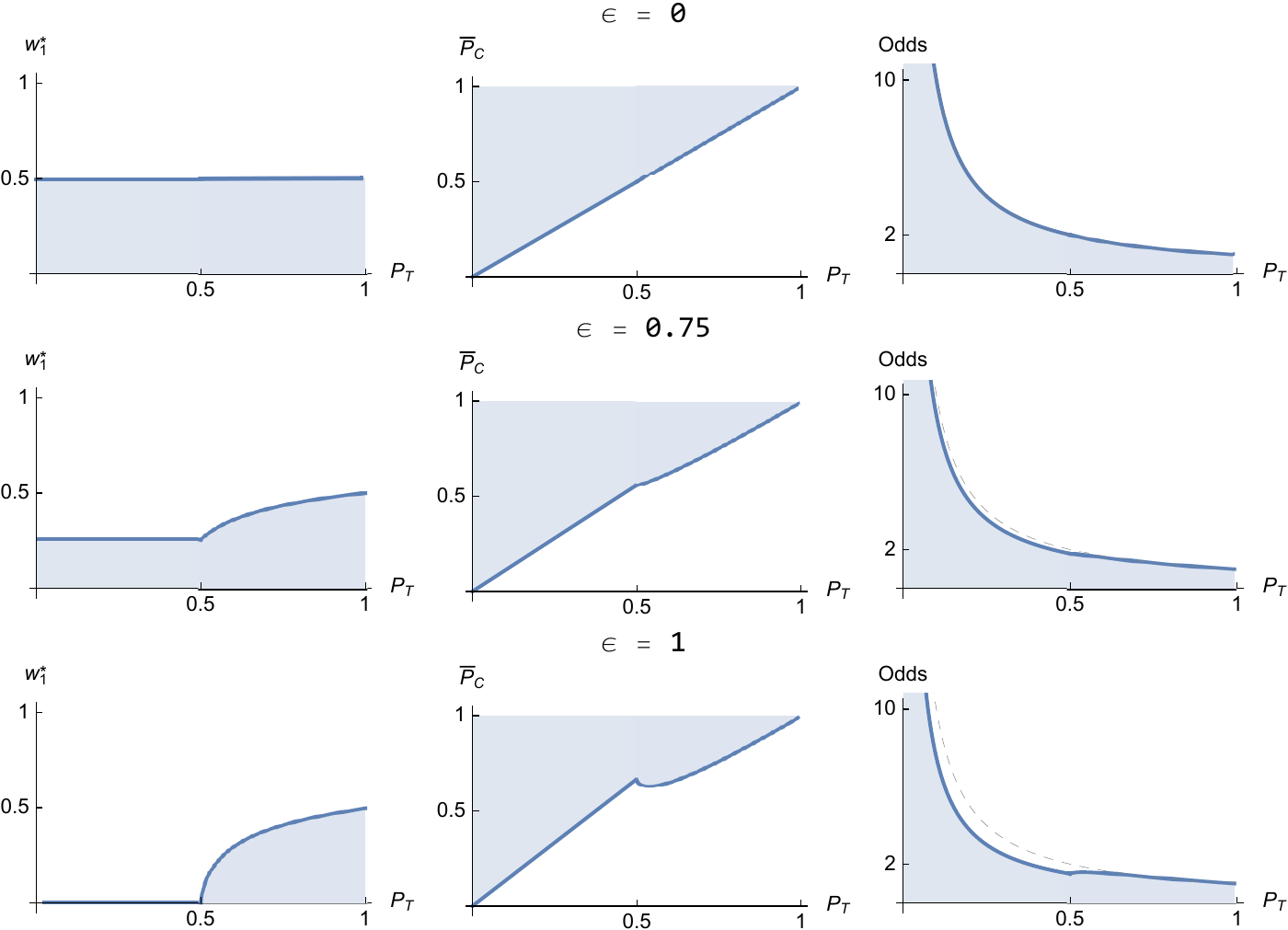}
    \caption{\normalfont The economic asymmetry that occurs between the buyer and the seller when they gamble indiscriminately at odds determined by an equal weighting of their noisy probabilities appears naturally countered and corrected by weighing their probabilities unequally. Specifically, the weight $w_1^*$ placed on the seller's probability across different relative frequencies $P_T$ (left) should generally be diminished, particularly for magnitudes of $P_T$ at the point of chance or below. Indeed, in the extreme case of $\epsilon = 1$, all weight should be laced on the buyer's probability. Only when $\epsilon = 0$ or $P_T = 1$ does the rule of equal weight provide symmetric economic conditions. If the market converges on this amendment to the basic game, then the predicted consequence would be odds that undervalue the chance occurrence and the longshot, but values the favorite broadly in line with $P_T$. We predict the market will reject this solution, however, because wagers that discriminate between odds under this solution have positive expectations.}
\end{figure}

\subsection*{Candidate B: Fair Odds for Noisy Probabilities}
Suppose you have tried to keep secret your invitations to play the basic game, but now find Bob on your trail. He has approached each of your other friends and asked about submitted probabilities, what probabilities were yours, and at what point you decided to abandon the game. Looking at the numbers, your strategy to quit the basic game whenever you have the role of the bookmaker becomes clear, and Bob notices just how well you have done overall; although you have lost bets, you are so far ahead that Bob must seriously consider that you have discovered some kind of edge.  At first, Bob merely decides he will congratulate you in public and thereby reveal your strategy. But then he starts looking at the numbers more closely and sees what appears to be a flaw in your behavior. While you were right to follow through on wagers whenever probabilities gave you the role of buyer - on the condition that odds were higher than two - you would have made more money if you abandoned the game whenever the odds were smaller than two. On the flip side, that means you were wrong to leave the game in the role of the bookmaker whenever odds were lower than two. Armed with this new insight, Bob draws up a simple 2 x 2 decision matrix (Figure 5) and heads toward the stadium. He hopes you will agree to his new terms, but not understand their importance and proceed as before.

\begin{figure}
    \centering
    \textbf{~~~~~~~~~Optimal Strategies for Players in the Basic Game:} \newline \par\medskip
\begin{tabular}{@{}lcc@{}}
         & $\frac{1}{P_C}>2.00$ & $\frac{1}{P_C}<2.00$ \\
         &                      & \multicolumn{1}{l}{} \\ \cmidrule(l){2-3} 
Buyer~~  & Bet                  & Abandon              \\ \cmidrule(l){2-3} 
Seller~~ & Abandon              & Bet                  \\ \cmidrule(l){2-3} 
         &                      & \multicolumn{1}{l}{}
\end{tabular}
\caption{\normalfont  When probabilities are noisy, and $P_T$ is distributed for a given magnitude of $P_C$, the seller can profit against an unwitting buyer by accepting wagers that have odds smaller than two, and rejecting bets that have odds greater than two. Conversely, the buyer can profit against an unwitting seller from wagering at odds greater than two and rejecting to wager at odds lower than two. Accordingly, the basic game must eventually come to an end unless gamblers agree to drive a wedge between the odds that are consistent with the relative frequencies of possible outcomes and the odds that are consistent with zero-expectation wagers.}
\end{figure}

What Bob has discovered is a crucial effect of noisy probabilities. $P_S$ and $P_B$ are distributed uniformly for a given magnitude of $P_T$, and $P_C$ consequently follows a triangular distribution with upper and lower limits given by $H$ and $L$ respectively, and with a mode given by solution \ref{eq:4}. But the corollary is that $P_T$ is distributed too, for each value of $P_C$ except the extremes. We can find the exact distribution of $P_T$ by assembling all the triangular distributions of $P_C$ across the interval $0 \leq P_T \leq 1$, and slicing, so to speak, the derived structure at the particular magnitude of $P_C$ in focus (Figure 6). Performing this procedure (see Appendix I) yields the probability density function 

\begin{equation}
\label{eq:18}
f(P_T) = {\begin {cases} \frac{(\epsilon -1) P_T + P_C}{\epsilon ^2 P_T^2} & : P_C\leq P_T\leq \frac{P_C}{1-\epsilon } \\ \frac{(\epsilon +1) P_T-P_C}{\epsilon ^2 P_T^2} & : \frac{P_C}{\epsilon +1}\leq P_T\leq P_C \\ 0 & : \text{Otherwise}. \end {cases}}\\
\end{equation}
for $2 \epsilon  P_T<\epsilon$, and

\begin{equation}
\label{eq:19}
f(P_T) = {\begin {cases} \frac{\epsilon - (\epsilon +1) P_T + P_C}{\epsilon ^2 \left(P_T-1\right){}^2} & : P_C\leq P_T\leq \frac{P_C+\epsilon }{\epsilon +1} \\ \frac{\epsilon +(1-\epsilon ) P_T -P_C}{\epsilon ^2 \left(P_T-1\right){}^2} & : \frac{\epsilon -P_C}{\epsilon -1}\leq P_T\leq P_C \\ 0 & : \text{Otherwise}. \end {cases}}\\
\end{equation}
for $2 \epsilon  P_T\geq \epsilon$.

\begin{figure}
    \centering
    \textbf{The Distribution of $P_T$ for Values of $P_C$:}\par\medskip
    \includegraphics[scale=0.45]{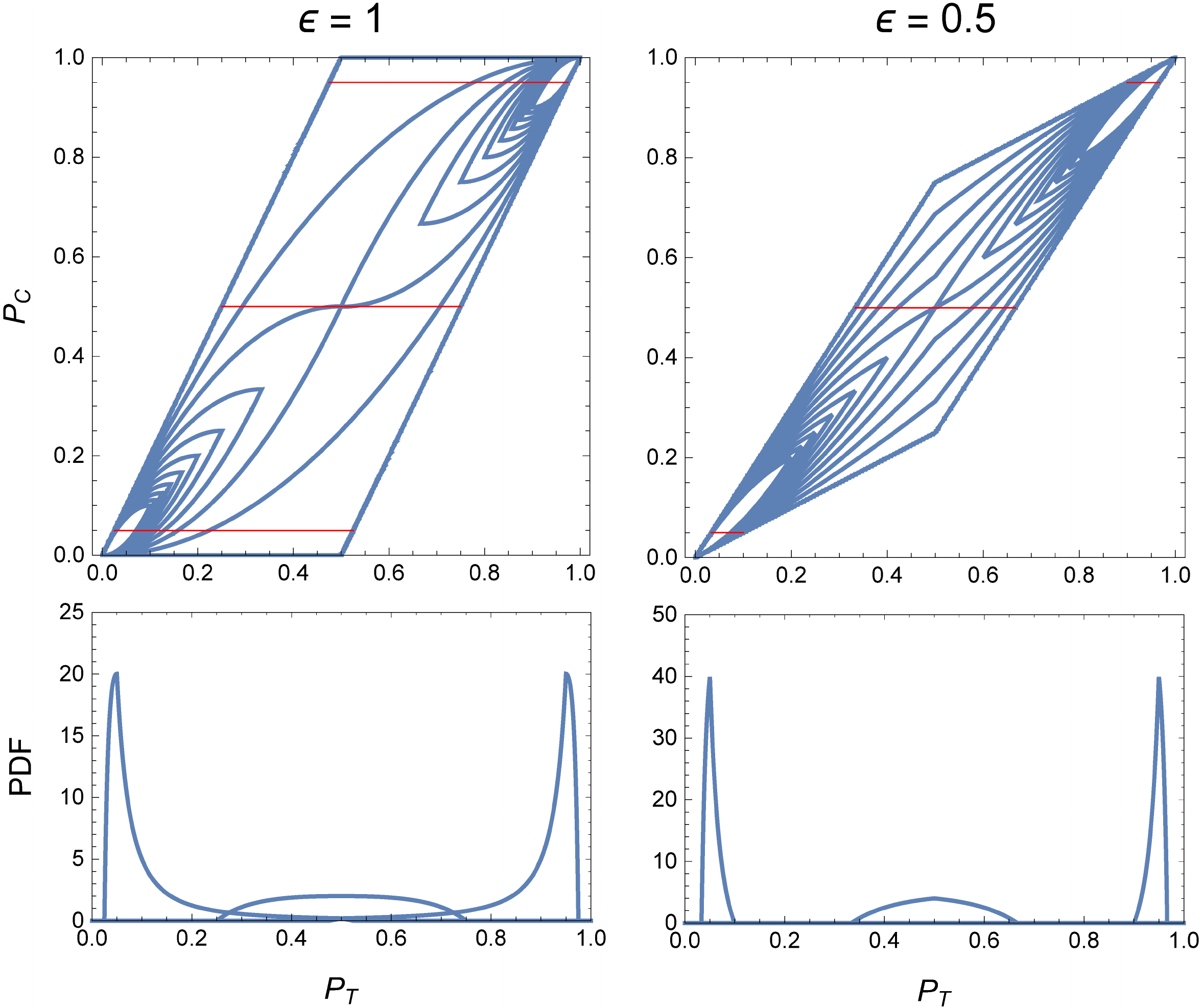}
    \caption{\normalfont The reason why the economic asymmetry in the basic game cannot simply be countered and corrected by giving different weight to different beliefs is due to another effect of noisy probabilities. Specifically, when $P_B$ and $P_S$ are distributed around a particular magnitude of $P_T$, then $P_T$ is distributed around a particular magnitude of $P_C$. Here we show the cases $\epsilon = 1$ (left) and $\epsilon = 0.5$, for $P_C = 0.05, P_C = 0.5$, and $P_C = 0.95$. As a consequence of $P_T$ being distributed around $P_C$, one observed magnitude of odds (i.e., $\frac{1}{P_C}$) will be associated with many different unobserved magnitudes of $P_T$, which creates profit opportunities for the buyer and the seller across different regions of odds.}
\end{figure}

Now, to answer if the seller's profit varies from negative to positive across the spectrum of $\frac{1}{P_C}$, as Bob suspects it does, we must calculate the mean of the seller's expected margin at particular values of $P_C$, when $P_T$ is distributed according to $f(P_T)$. In doing so, however, we prepare for our final and most important point by introducing the constant $m$ to capture by how much $P_C$ must increase or decrease at each magnitude of $P_C$ to yield zero-expectation bets across the entire spectrum, and thereby keep the game alive. 

Appendix I contains the precise mathematical expressions of $\bar{\pi }_{S_{o}}$ under different settings of $\epsilon$ and $P_C$, while the left side of Figure 7 visually confirms what Bob guessed: for values of $\frac{1}{P_C} < 2$, the seller's mean expected margin is indeed positive, while for values of $\frac{1}{P_C} > 2$, the seller's mean expected margin is negative. 

\begin{figure}
    \centering
    \textbf{Candidate Solution B: Fair Odds for Noisy Probabilities:}\par\medskip
    \includegraphics[scale=1]{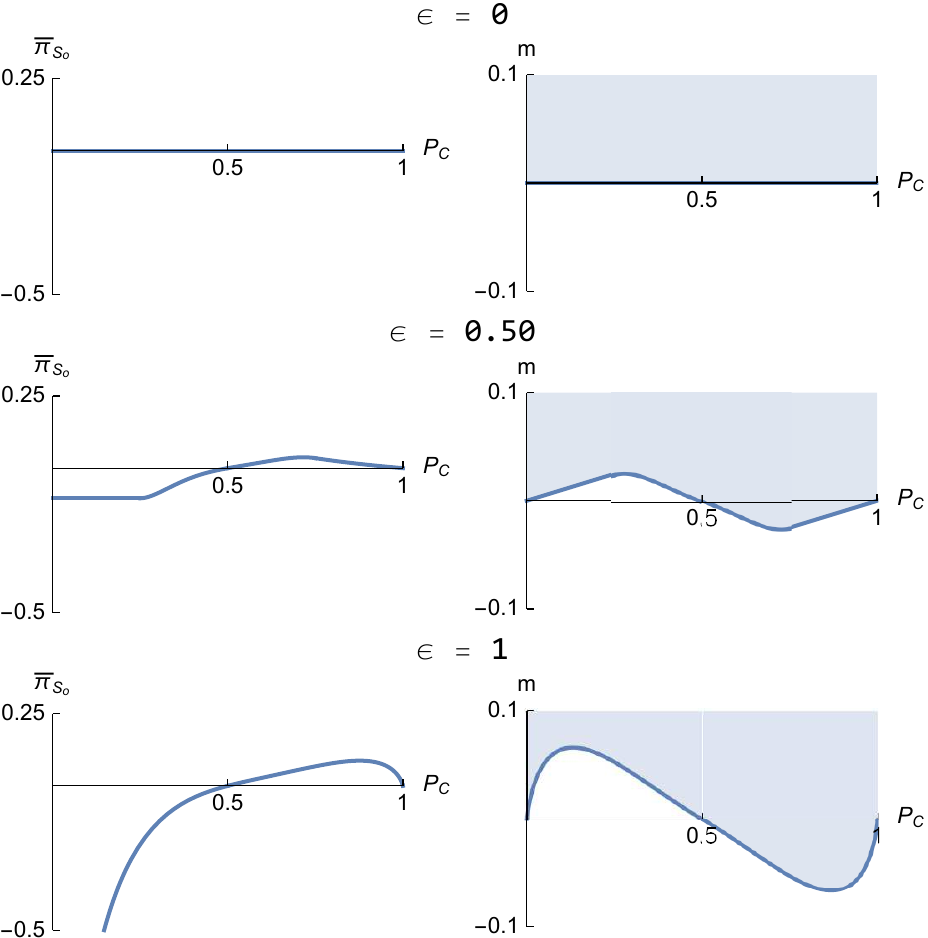}
    \caption{\normalfont Since the seller loses money in the basic game by wagering at odds greater than two, and the buyer loses cash by wagering at odds smaller than two (left), players must reduce odds for unlikely events (i.e., $m$ must be added to $P_C$ for $P_C < 0.5$), and increase odds for likely events ($m$ must be deducted from $P_C$) to make odds fair. As illustrated on the right, the amount by which values of $P_C$ below $0.5$ must be increased corresponds precisely to the amount by which the corresponding magnitude at $1 - P_C$ must be decreased, thus keeping the sum of $P_C$'s for the same binary option equal to one. Odds in the basic game are fair without adjustment when probabilities are noiseless.}
\end{figure}

Thus, although someone might agree to Bob's new terms initially, they would find their bankroll dwindle if only Bob applied the change, and  given that realization, they would stop participating.

For the game to continue indefinitely, what is required is an understanding between players that odds for the favorite ($\frac{1}{P_C} > 2$) must be increased above those given by solution \ref{eq:4}, while they must decrease for the underdog ($\frac{1}{P_C} < 2$). By how much $P_C$ must be increased or decreased to keep the game alive is found by first setting $\bar{\pi }_{S_{o}}$ to zero and solving the derived equations for $m$. Appendix I shows these details mathematically, while Figure 8 and Figure 9 illustrate them visually. In short, gambling based on noisy probabilities is predicted to cause an adjustment to the basic game whereby odds undervalue the favorite, and overvalue the long shot; although misaligned with the relative frequency of outcomes at the limit, the odds are now fair given the level of noise.

\begin{figure}
    \centering
    \textbf{Fair Odds for Noisy Probabilities Compared with $\frac{1}{P_T}$:}\par\medskip
    \includegraphics[scale=1]{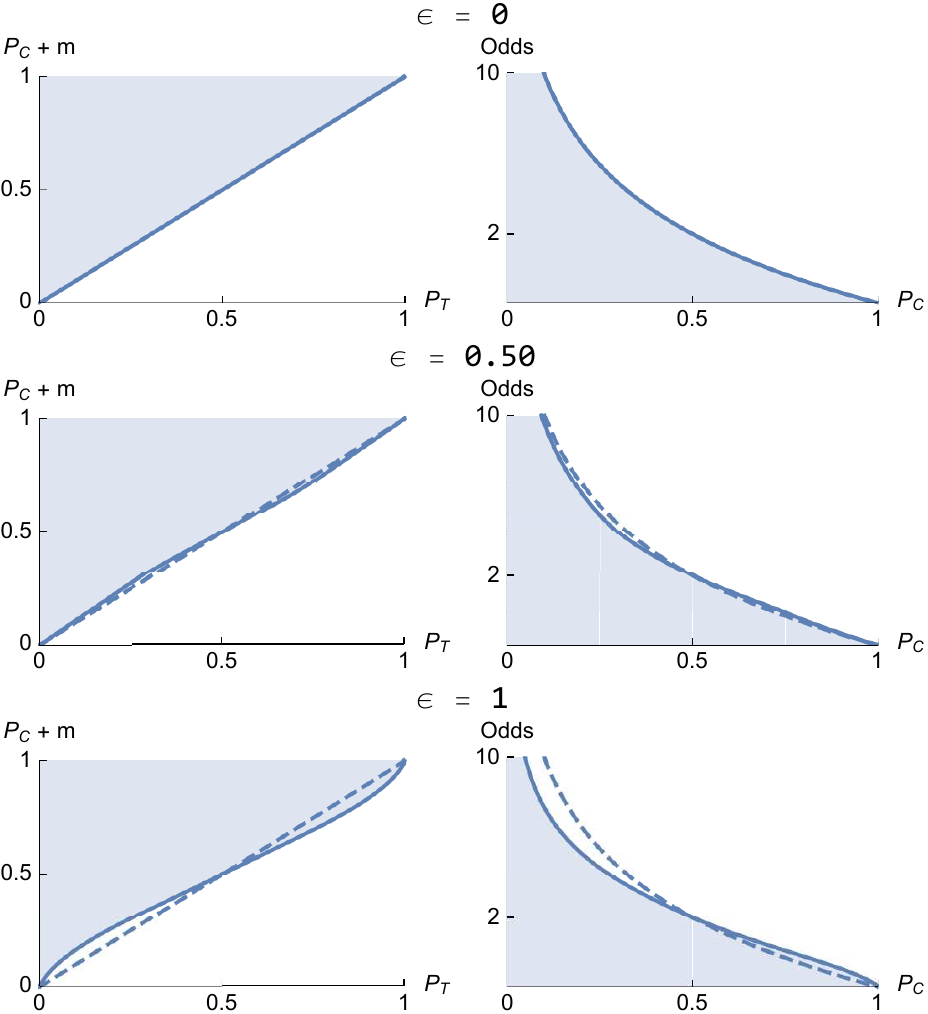}
    \caption{\normalfont To achieve fair odds for noisy probabilities, that is to say, odds associated with zero-expectation wagers, gamblers must agree to overvalue the longshot and undervalue the favorite.  In other words, they must drive a wedge systematically between odds associated with zero-expectation wagers and odds consistent with the relative frequencies. The amount by which odds must under and overvalue $P_T$ depends on the level of noise in the buyer's and the seller's probabilities.}
\end{figure}

\pagebreak

\section*{Discussion}     
In the face of uncertainty, people gamble \citep{FrankPRamsey1926, Reichenbach1938a, Brunswik1943a}, but they also gather evidence for and against alternative hypotheses to improve their degrees of belief. Economists \citep{Stigler1961, Arrow1973} have long recognized that people invest resources gathering information while trying to manage the trade-off between decision speed and accuracy. Less studied, however, is 
the fact that people display variability in the choices they make when confronted again with the same evidence \citep{Faisal2008}. 

One of the candidate theories for such choice variability involves the noisy accumulation of evidence by competing populations of neurons for and against alternative views about the most significant source of reward \citep{Yang2007, Ratcliff2016a, Pisauro2017, SteingroeverHelen2018}.  Indeed, there is growing support for the idea that neurons in these populations regulate their activity in striking accordance with the weight of evidence \citep{Good1985a, Gold2002a, Gold2007a}. That is intriguing because it suggests the brain has evolved to gather evidence in ways remarkably similar to sequential analysis (SA), invented by Turing to help England break Enigma during World War II, and by Wald to determine if batches of munitions were of satisfactory quality to ship by the US Army during the same conflict.

In SA, new evidence and updated probabilities follow each other deterministically through Bayes' factor in the string of samples. Moreover, different sequences through the body of evidence ultimately lead to the same posterior probability distribution. But in the neurophysiological counterpart, noise between the presented evidence and its internal representation by neurons arguably makes the posterior probability distribution different from the one produced by the precision of mathematics. While SA computes one probability for each possible outcome, given the sampled evidence, it would appear the neurophysiological counterpart produces not one probability for each possible outcome, but an entire distribution of probabilities, given the same evidence. Accordingly, probability itself becomes a random variable. 

We examined the economic consequence of such "noisy probabilities" for odds agreed between a buyer and a seller of binary options. Noisy, but unbiased probabilities were predicted to compel these gamblers to drive a wedge between the odds that are consistent with the relative frequencies of alternative outcomes, and the odds that are consistent with zero-expectation bets. In particular, the buyer and the seller agreed to odds that were higher than the reciprocal of their averaged unbiased beliefs when this average indicated the outcome was more plausible than chance, while they agreed to odds that were lower than the reciprocal of their averaged unbiased beliefs when this average suggested the result was less likely than chance. The gamblers had to do this to cancel an unfairness towards the seller in odds greater than two, and cancel an inequity towards the buyer in odds smaller than two, which exist when probabilities are noisy, and gamblers place equal weight on their respective beliefs.

Accordingly, our predictions suggest not only that the definition of fair odds must account for the level of noise in the degrees of belief held by those who agree on odds, but also that noisy probabilities could play an essential role in creating the long-standing empirical regularity of the favorite-longshot bias \citep{Griffith1949, Shin1992, Ottaviani2010a, Snowberg2010}. Perhaps of equal interest is that our work suggests noisy probabilities make the old-established way of revealing degrees of belief through wagering more difficult than previously thought \citep{FrankPRamsey1926,DeFinetti1937,Carnap1962,Edwards1963}, and thereby casts doubt on the promise \citep{Arrow} of betting markets to provide unbiased indicators of future events and support rational decisions.

\subsection*{The Fairness of a Favourite-Longshot Bias} Since \citet{Griffith1949} first noted the importance for psychology and economics of understanding why betting markets tend to undervalue the favorite and overvalue the longshot, researchers in both fields have proposed numerous theories about why the phenomenon occurs. We can suitably categorize current theories about the favorite-longshot bias either as conjectures about the effect of individual cognitive bias that individuals bring to the market or as conjectures about the effect of social interaction in the market. In the first category, we find what \citet{Snowberg2010} call the risk-love and misperception of probability explanations, while in the second category we find reasons that revolve around the presence of private information. 

The risk-love explanation is the neoclassic point of view, asserting that gamblers are rational but have utility functions with preferences for risk \citep{Friedman1952}. Accordingly, gamblers wager high volumes on the longshot compared to the favorite, causing odds on the longshot to drop below the level that reflects the relative frequency of winning, while odds on the favorite move in the other direction. In contrast, the misperception of probability explanation is the behavioral point of view, asserting that gamblers neither intend to overvalue the longshot, nor undervalue the favorite, but do both due to cognitive errors that create biased judgments. Laboratory studies showing that people have difficulty discriminating between different magnitudes of small probabilities support this theory and has resulted in the development of the probability-weighting-function \citep{Prelec1998}, which reflects the bias observed in markets.

Among explanations that involve private information, we find the idea \citep{Shin1992} that bookmakers have customers with insider information whom they cannot distinguish from customers with only public information. Bookmakers are, however, assumed to know the proportion of customers with unique information, and their optimal defense under these circumstances is to reduce odds on the longshot and increase odds on the favorite according to the fraction of insiders they face. Meanwhile, another theory \citep{Ottaviani2010a} in this category asserts that when collectives of people in possession of informative private signals wager simultaneously, they will be surprised to find that although favorites are indeed more likely to win, as their signal suggests, favorites win even more often than what the transacted odds imply, while longshots win less often. In contrast, when signals are uninformative (the authors of the article talk of noisy signals, and these must not be confused with noisy probabilities) the favorite-longshot bias disappears, or reverses.  

Our predictions link to cognitive phenomena, just like theories in the first described category, but do not involve cognitive bias. Moreover, like Shin's theory in the second category, our predictions are based on adjustments to cancel the prospect of economic inequity but do not involve unequal access to information. Instead, we show that when gamblers accumulate evidence about uncertain events stochastically, and when their degrees of belief consequently couple randomly to the whole evidence, odds that reflect the relative frequencies of outcomes are inequitable and must be adjusted to establish zero-expectation wagers. Accordingly, we suggest the favorite-longshot bias is an essential response by the market to create fairness among participants.

We are not suggesting the effect of noisy probabilities works alone; we acknowledge that empirical studies show that gambling on the long-shot results in worse performance than gambling on the favorite. However, this observation does not rule out the predicted effect of noisy probabilities. Instead, it may be that preferences for risk work together with noisy probabilities to make the favorite-longshot bias more pronounced, creating demand and supply conditions that move odds away from their zero-expectation levels.
 
\subsection*{The Unkept Promise of Betting Markets} An important debate among economists concerns how we should interpret the reciprocal of odds. One argument \citep{Wolfers2006b} maintains not only that we should understand this reciprocal as the market's degree of belief, but also that such beliefs by markets are consistent with the mean belief across market participants, and tends to be the best predictor of the event in question. As such, the argument summons the rational expectations hypothesis stated by \citet{Muth1961}, who argued that distributions of subjective probabilities tend to scatter about the prediction of the relevant theory. On the other hand, the counter-argument \citep{Manski2006} asserts that odds reflect the mean belief of participants only under particular circumstances, and points out that systematic undervaluation of favorites and overvaluation of longshots are anomalies inconsistent with the efficiency depicted by proponents. 

The debate has more than just implication for the efficient market hypothesis because the truth of the matter affects the possibility of harnessing the wisdom of crowds through betting to guide well-informed decisions. If skeptics are right, the promise of betting markets to produce forecasts of outcomes with a lower error than most conventional forecasting methods \citep{Arrow}, cannot generally be kept. 

Our predictions support the skeptics by suggesting that odds must systematically deviate from the mean beliefs of market participants to be fair. However, as can readily be appreciated without much handwaving, on the general assumption that beliefs closest to, but on either side of the median determine odds, the effective value of $\epsilon$ approaches zero as the number of gamblers increases. Moreover, when the effective range of noisy probability thereby closes, it does so at $P_T$. Nevertheless, for any market of limited size, the effect is insufficient to eradicate the bias.

\subsection*{The Problem of Revealing Probabilities} We must necessarily end on the most fundamental note because the presence of noisy probabilities adds some complication to the old-established way of revealing someone's degree of belief, which has been used regularly as a thought experiment to explain the Bayesian position \citep{FrankPRamsey1926, DeFinetti1937, Edwards1963}. In the classic thought-experiment \citep{DeFinetti1937}, someone is trying to expose the degree of belief held by someone else through the proposal of a wager (Gilles (2000) calls them Ms. A and Mr. B respectively, and we use those names here). In particular, Ms. A asks Mr. B to state the odds at which he is willing to risk losing a stake that Ms. A can decide, on the further condition that once Mr. B has revealed these odds, Ms. A is also allowed to determine what side of the market she prefers. By applying this scheme, Ms. A \textbf{appears} to create a position where she can force an honest answer from Mr. B because Ms. A can punish Mr. B instantly through her choice of role, should Mr. B be dishonest in an attempt to gain an economic advantage. However, that appearance misleads when probabilities are noisy because under these conditions the game punishes honesty, not dishonesty.

By allowing Ms. A to choose what side of the market she prefers after Mr. B has set the odds, Mr. B is essentially letting Ms. A wager at odds that place no weight on her belief, and consequently, Ms. A can at the very least avoid expected losses and at best secure expected profits. Although our previous story about Bob assumed equal weighting of beliefs, the optimal strategies for Bob and Ms. A, and hence our conclusions, are the same. Mr. B must adjust his odds downward when he believes the probability of winning is smaller than chance and move his odds upward when he thinks the probability of victory is higher than chance. Indeed, if he calibrates this adjustment to precision, he should not care which side of the market Ms. A ends up choosing (see Appendix II). Of course, the odds that Mr. B reveals to Ms. A will not reflect what he thinks, and they will not be consistent with the relative frequencies of the given situation. However, they will be fair.

\pagebreak

\appendix\setcounter{secnumdepth}{0}

\section{Appendix I}
In this appendix, we derive the probability density function for $P_T$, $f(P_T)$, for various odds $\frac{1}{P_C}$ agreed by players of the basic game when their personal probabilities are noisy. We then use $f(P_T)$ to derive the mean expected margin $\pi _{S_o}$ of these players in their role of the seller, and find the degree $m$ by which $P_C$ must be adjusted to yield zero-expectation wagers. Since the favorite-longshot bias emerges at the end of this procedure, there is reason to suspect this long-standing empirical phenomenon involves an adjustment to establish the foundation of an equitable market.

Given our assumptions about zero market power, risk neutrality, and unbiased uniform noise $\epsilon$, and because we assume players of the basic game place equal weight on their respective degrees of belief, the consensus belief $P_C$ about $P_T$ follows the triangular distribution. More specifically, $P_C$ follows the triangular distribution with a lower limit given by $L$ and an upper limit given by $H$ from equation \ref{eq:5}, while its mode is given by solution \ref{eq:4}, which makes the distribution symmetric. As a corollary of this arrangement, there is a distribution of $P_T$ for any magnitude of $P_C$, and our initial task is to find that distribution. 

Our procedure starts with the observation that $L$ and $H$ are governed by the circumference of the rhombus depicted in Figure I. In particular, the left and top of the rhombus determine $H$, while $L$ is determined by the right and bottom. We capture this using the following equations:

\begin{equation*}
\begin{aligned}
\text{left} &= (1 + \epsilon) P_T \\
\text{top} &= (1 - \epsilon) P_T + \epsilon \\
\text{right} &= (1 + \epsilon) P_T - \epsilon \\
\text{bottom} &= (1 - \epsilon) P_T \\
H &= \min (\text{left},\text{top}) \\
L &= \max (\text{right},\text{bottom}) \\
\end{aligned}
\end{equation*}

Since the probability density function of a symmetric triangular distribution is given by

\begin{equation*}
\label{eq:AI1}
{\begin {cases} 

\frac{4 (b-x)}{(b-a)^2} & : a+b<2 x\land b\geq x \\ 

-\frac{4 (a-x)}{(b-a)^2} & : a\leq x\land a+b\geq 2 x \\ 

0 & : Otherwise \\ 

\end {cases}}\\
\end{equation*}
where $a$ is the lower limit and $b$ is the upper limit, we substitute $L = \max (\text{right},\text{bottom})$ for $a$ and $H = \min (\text{left},\text{top})$ for $b$ to find the probability density function for $P_T$ given by equations \ref{eq:18} and \ref{eq:19} in the main text. These equations allow us now to hold $\frac{1}{P_C}$ constant and investigate how $P_T$ scatters when players wager at these odds. 

Accordingly, we ask what margins players can expect at particular odds set in accordance with solution \ref{eq:4} in the role of seller when $P_T$ is distributed by $f(P_T)$. The margin of players in the role of seller is given by equation \ref{eq:3} except we substitute $P_S$ with $P_T$, thus changing the margin from subjective $\pi_{S_s}$ to objective $\pi_{S_o}$. The answer to our question is the expectation of $\pi_{S_o}$, but since out main purpose is to discover the degree $m$ by which $P_C$ must be adjusted to result in zero-expectation wagers, we find the expectation of

\begin{equation*}
\pi _{S_o}=\left(1-P_T\right)-P_T \left(\frac{1}{P_C+m}-1\right),~P_T \sim f(P_T).
\end{equation*}

The mean of the seller's expected margin $\bar{\pi }_{S_{o}}$, which the left side of Figure VII visualizes, consists of different segments to cover the range of $P_C$. Some of the segments are rather lengthy, but a series of sub-expressions appear regularly across these segments, and we can simplify things markedly by substituting these sub-expressions for variables as done below

\begin{equation*}
\begin{aligned}
x_1 &= \ln (a+1)\\
x_2 &= \ln (1-a)\\
x_3 &= \ln \left(1-P_C\right)\\
x_4 &= \ln \left(\frac{1}{P_C}\right) \\
x_5 &= \ln \left(\frac{1}{P_C^2}\right) \\
x_6 &= \ln \left(\frac{1-P_C}{P}\right) \\
x_7 &= \ln \left(\frac{-a-1}{-P_C-1}\right)\\
x_8 &= \ln \left(\frac{1-P_C}{1-a}\right)\\
x_9 &= \ln \left(P_C\right) \\
x_{10} &= \ln \left(P_C^4\right) \\
x_{11} &= \ln (2) \\
x_{12} &= \ln \left((a+1) \left(-\left(1-P_C\right)\right)\right) \\
x_{13} &= \tanh ^{-1}\left(2 P_C-1\right) \\
x_{14} &= 2 x_1-x_3-x_9-2 x_{11}-2 \\
x_{15} &= 2 x_3+2 x_{11}+x_{14} \\
x_{16} &= -x_3-x_{11}-x_{14}-2 \\
x_{17} &= x_{14} \left(P_C+m\right)+x_{11} \\
\end{aligned},
\end{equation*}

After performing these substitutions, the mean of the seller's expected margin is given by the following system of piecewise equations

\begin{equation*}
\label{eq:AI1}
\bar{\pi }_{S_{o}} = {\begin {cases} 

\frac{\left((\epsilon+2) x_1-(\epsilon-2) x_2\right) P_C+m \left((\epsilon+1) x_1-(\epsilon-1) x_2\right)}{\epsilon^2 \left(P_C+m\right)} & : \text{i} \\ 

\frac{\epsilon \left(x_1-x_2\right) \left(P_C+m-1\right)+\left(x_1+x_2\right) \left(2 P_C+m-2\right)}{\epsilon^2 \left(P_C+m\right)} & : \text{ii} \\ 

\frac{2 m \left((\epsilon+1) x_1-\epsilon\right)}{\epsilon^2 \left(P_C+m\right)} & : \text{iii} \\ 

\frac{2 m \left(x_3 \left(-P_C\right)+x_9 P_C+x_3\right)+2 x_{11} \left(2 P_C+m-2\right)-P \left(2 x_6 P_C-5 x_3+x_9\right)-3 x_3+\frac{x_5}{2}}{\epsilon^2 (m+P)} & : \text{iv} \\ 

\frac{P_C \left(-2 P_C-3 x_3+3 x_{11}+1\right)-m \left(2 P_C+2 x_3-2 x_{11}+1\right)+3 x_3}{P_C+m} & : \text{v} \\  

\frac{m \left(2 P_C+2 x_4+2 x_{11}-3\right)+P_C \left(2 P_C+3 x_4+3 x_{11}-3\right)-3 x_{11}+1}{P_C+m} & : \text{vi} \\ 

\frac{\epsilon \left(-x_7+x_{17}+1\right)+4 P_C \left(P_C+m+x_1-x_{13}-1\right)+m x_{15}-2 x_{11}-2 x_{12}+1}{\epsilon^2 \left(P_C+m\right)} & : \text{vii} \\ 

\frac{2 m x_{16} P_C+x_1 \left(2 m \left(P_C+1\right)+P_C \left(5-2 P_C\right)\right)+\frac{1}{2} \left(x_{10}+4 x_{11}\right) P_C^2+x_{16} P_C}{\epsilon^2 \left(P_C+m\right)} & : \text{viii} \\ 

\frac{\epsilon \left(x_8+x_{17}+1\right)+4 P_C \left(-P_C-m+x_1+x_{13}+1\right)+4 m x_1-m x_{15}+2 x_8+2 x_{11}-1}{\epsilon^2 \left(P_C+m\right)} & : \text{ix} \\ 

\end {cases}}\\
\end{equation*}
\linebreak
where the Roman numerals denote the conditions for which the corresponding piece of the overall system is relevant. These are conditions are  

\footnotesize
\begin{equation*}
\label{eq:AI2}
\text{i to iv : } {\begin {cases} 
\epsilon +1\geq 2 P_C\land 0<P_C\leq \frac{1}{2}\land \left(\left(0<\epsilon <\frac{1}{2}\land 2 P_C+\epsilon \leq 1\right)\lor \left(\frac{1}{2}\leq \epsilon <1\land 2 P_C+\epsilon <1\right)\right) \\ 

\frac{1}{2}<P_C<1\land \epsilon +1<2 P_C\land \left(\left(2 P_C+\epsilon >1\land \epsilon >0\right)\lor \left(2 P_C+\epsilon \geq 1\land 2 \epsilon \geq 1\right)\right) \\ 

2 P_C=1\land \epsilon <1\land \epsilon >0 \\ 

\epsilon +1=2 P_C\land \frac{1}{2}<P_C<1\land \epsilon <1 \\ 

\epsilon =1\land 0<P_C\leq \frac{1}{2} \\  

\epsilon =1\land 2 P_C>1\land P_C<1  \\ 

\epsilon +1>2 P_C\land 2 P_C>1\land \epsilon <1\\ 

\frac{1}{2}\leq \epsilon <1\land 0<P_C<\frac{1}{2}\land 2 P_C+\epsilon =1 \\ 

\epsilon <1\land 2 P_C<1\land 2 P_C+\epsilon >1 \\ 

\end {cases}}\\
\end{equation*}\
\normalsize
\linebreak

\noindent Finally, the equations for the magnitude by which $P_C$ must be adjusted to ensure fair odds $\frac{1}{P_C + m}$ are found by setting each of the piecewise equations above to zero and then solving for $m$. Performing these calculations yields 

\begin{equation*}
\label{eq:AI3}
m = {\begin {cases} 

\frac{\left((\epsilon-2) x_2-(\epsilon+2) x_1\right) P_C}{(\epsilon+1) x_1-(\epsilon-1) x_2} & : \text{i} \\ 

\frac{\left((\epsilon-2) x_2-(\epsilon+2) x_1\right) \left(P_C-1\right)}{(\epsilon+1) x_1-(\epsilon-1) x_2} & : \text{ii} \\ 

0 & : \text{iii} \\ 

\frac{P_C \left(-4 x_6 P_C+10 x_3-2 x_9\right)+8 x_{11} \left(P_C-1\right)-6 x_3+x_5}{8 x_{13} P_C+4 x_3+4 x_{11}} & : \text{iv} \\ 

\frac{P_C \left(-2 P_C-3 x_3+3 x_{11}+1\right)+3 x_3}{2 P_C+2 x_3-2 x_{11}+1} & : \text{v} \\  

\frac{P_C \left(-2 P_C-3 x_4-3 x_{11}+3\right)+3 x_{11}-1}{2 P_C+2 x_4+2 x_{11}-3} & : \text{vi} \\ 

\frac{\epsilon \left(x_7-x_{11}-1\right)+\left(-2 x_1+2 x_{13}+2\right) P_C+2 x_{11}+2 x_{12}-1}{\epsilon \left(2 x_1-x_3-x_9-2 x_{11}-2\right)+4 P_C+2 x_1-2 x_{13}-2}-P_C & : \text{vii} \\ 

\frac{P_C \left(x_1 \left(6-4 P_C\right)+\left(x_{10}+4 x_{11}\right) P_C+2 \left(x_9+x_{11}\right)\right)}{4 x_1 \left(P_C-1\right)-4 \left(x_9+x_{11}\right) P_C} & : \text{viii} \\ 

\frac{\epsilon \left(x_8+x_{11}+1\right)+\left(2 x_1+2 x_{13}+2\right) P_C+2 x_8+2 x_{11}-1}{\epsilon \left(-2 x_1+x_3+x_9+2 x_{11}+2\right)+4 P_C-2 x_1-2 x_{13}-2}-P_C & : \text{ix} \\ 
\end {cases}}\\
\end{equation*}
\linebreak
\section{Appendix II}
Here we show the equations for the seller's mean expected margin in De Finetti's classic thought experiment when honest probabilities are drawn from the uniform distribution described by equation \ref{eq:5} in the main text. We then show the equations for the magnitude by which $P_C$ must be adjusted to establish odds that are fair. These adjustments resemble those required when $P_C$ is determined by weighing $P_S$ and $P_B$ equally as in the basic game. 

Once again, our initial task is to assemble distributions of $P_C$ across the interval $0 \leq P_T \leq 1$, and slice the derived structure at the particular magnitude of $P_C$ to find the distribution of $P_T$ at this point. However, unlike before, in De Finetti's thought experiment $P_C$ is derived by placing all the weight on the subject's honest probability, such that the distribution of $P_C$ is given by equation \ref{eq:5}. Nevertheless, the basic procedure resembles the one described in Appendix I. 

Accordingly, we start with the observation that $L$ and $H$ are governed by the circumference of the rhombus depicted in Figure I and again arrive at the following equations

\begin{equation*}
\begin{aligned}
\text{left} &= (1 + \epsilon) P_T \\
\text{top} &= (1 - \epsilon) P_T + \epsilon \\
\text{right} &= (1 + \epsilon) P_T - \epsilon \\
\text{bottom} &= (1 - \epsilon) P_T \\
H &= \min (\text{left},\text{top}) \\
L &= \max (\text{right},\text{bottom}) \\
\end{aligned}
\end{equation*}

This time, however, we must substitute $H = \min (\text{left},\text{top})$ and $L = \max (\text{right},\text{bottom})$ into the probability density function of the uniform distribution

\begin{equation*}
\label{eq:AI1}
{\begin {cases} 

\frac{1}{b-a} & : a\leq x\leq b \\ 

0 & : Otherwise \\ 

\end {cases}}\\
\end{equation*}
where $a$ is the lower limit and $b$ is the upper limit of the distribution. Substituting $L = \max (\text{right},\text{bottom})$ for $a$ and $H = \min (\text{left},\text{top})$ for $b$ yields the probability the probability density function for $P_T$

\footnotesize
\begin{equation*}
\label{eq:AII1}
f(P_T) = {\begin {cases} \frac{1}{2 \epsilon (1-P_T)} & : P_T-\max \left(\frac{\epsilon-P_C}{\epsilon -1},\frac{P_C}{\epsilon +1}\right)\geq 0\land x-\min \left(\frac{-P_C}{\epsilon -1},\frac{P_C+\epsilon}{\epsilon +1}\right)\leq 0\land 2 \epsilon  P_T-\epsilon \geq 0 \\ 

\frac{1}{2 \epsilon  P_T} & : P_T-\max \left(\frac{\epsilon-P_C}{\epsilon -1},\frac{P_C}{\epsilon +1}\right)\geq 0\land x-\min \left(\frac{-P_C}{\epsilon -1},\frac{P_C+\epsilon}{\epsilon +1}\right)\leq 0 \\ 

0 & : \text{Otherwise}. \end {cases}}\\
\end{equation*}
\normalsize

From here, the procedure described in Appendix I can be followed exactly; we must derive the mean of the seller's expected margin $\left(1-P_T\right)-P_T \left(\frac{1}{P_C+m}-1\right)$ when $P_T$ is distributed according to $f(P_T)$, introducing $m$ in preparation for the final step to by how much $P_C$ must be adjusted to establish fair odds. Again $\bar{\pi }_{S_{o}}$ consists of different segments to cover the range of $P_C$, while a series of sub-expressions appear regularly across these:

\begin{equation*}
\begin{aligned}
x_1 &= x_1=\ln \left(\frac{1-P_C}{\epsilon +1}\right)\\
x_2 &= \ln \left(\frac{1-P}{P_C}\right)\\
x_3 &= \ln \left(\frac{P_C}{\epsilon +1}\right)\\
x_4 &= \ln \left(\frac{\epsilon +1}{P_C}\right) \\
x_5 &= \ln (2) \\
x_6 &= \tanh ^{-1}(\epsilon ) \\
x_7 &= -x_1-x_3-2 x_5\\
x_8 &= 4 x_5-4 x_4\\
\end{aligned}
\end{equation*}

After performing the required substitutions, the mean of the seller's expected margin is given by the following system of piecewise equations
\begin{equation*}
\label{eq:AII2}
\bar{\pi }_{S_{o}} = {\begin {cases} 

\frac{\frac{P_C}{\left(\epsilon^2-1\right)}}{{(m+P_C)}}+\frac{x_6}{\epsilon} & : \text{i} \\ 

\frac{\frac{P_C-1}{\epsilon^2-1}+\frac{x_6 (m+P_C-1)}{\epsilon}}{m+P_C} & : \text{ii}   \\ 

\frac{2 P_C x_2 (-m-P_C+1)+2 P_C-1}{4 \epsilon P_C (m+P_C)}& : \text{iii}  \\ 

\frac{(\epsilon+1) x_7 (m+P_C)+(\epsilon+1) x_1+(\epsilon+1) x_5+2 P_C-1}{2 \epsilon (\epsilon+1) (m+P_C)} &: \text{iv} \\ 

\frac{-\epsilon+(P_C-1) x_8 (m+P_C)+2 P_C-1}{4 \epsilon (\epsilon+1) (m+P_C)} &: \text{v} \\ 

0 & : \text{Otherwise} \\

\end {cases}}
\end{equation*}
where the Roman numerals again denote the conditions for which the corresponding piece of the overall equation is relevant. These are  

\footnotesize
\begin{equation*}
\label{eq:AII3}
\text{i to v} : {\begin {cases} 

\epsilon +1\geq 2 P_C\land \left(\left(0<\epsilon <\frac{1}{3}\land \left(2 P_C+\epsilon =1\lor \left(P_C>0\land 2 P_C+\epsilon \leq 1\right)\right)\right)\lor \left(P_C>0\land 3 \epsilon \geq 1\land 2 P_C+\epsilon <1\right)\right)\\ 

\epsilon +1<2 P_C\land P_C<1\land \left(\left(2 P_C+\epsilon >1\land \epsilon >0\right)\lor \left(2 P_C+\epsilon \geq 1\land 3 \epsilon \geq 1\right)\right)\\ 

2 P_C+\epsilon >1\land \epsilon <1\land \epsilon +1=2 P_C\\ 

\epsilon <1\land \epsilon +1>2 P_C\land 2 P_C+\epsilon >1 \\ 

\frac{1}{3}\leq \epsilon <1\land \epsilon +1>2 P_C\land 2 P_C+\epsilon =1\\  

\end {cases}}\\
\end{equation*}
\normalsize
\linebreak

The equations for the magnitude by which $P_C$ must be adjusted to ensure fair odds $\frac{1}{P_C + m}$ are now found by setting each of the above piecewise equations to zero and solving for $m$. This gives

\begin{equation*}
\label{eq:A4}
m = {\begin {cases} 
\frac{P_C \left(\left(\epsilon^2-1\right) x_6+\epsilon\right)}{\left(1-\epsilon^2\right) x_6} & : \text{i} \\ 

\frac{(P_C-1) \left(\left(\epsilon^2-1\right) x_6+\epsilon\right)}{\left(1-\epsilon^2\right) x_6} & : \text{ii}   \\ 

\frac{2 P_C-1}{2 P_C x_2}-P_C+1 & : \text{iii}  \\ 

-\frac{P_C ((\epsilon+1) x_7+2)+(\epsilon+1) x_1+(\epsilon+1) x_5-1}{(\epsilon+1) x_7} &: \text{iv} \\ 

\frac{\epsilon-2 P_C+1}{(P_C-1) x_8}-P_C &: \text{v} \\ 

0 & : \text{Otherwise}
\end {cases}}
\end{equation*}

As Figure IX illustrates, the adjustments of $P_C$ needed to establish fair odds in De Finetti's thought experiment resemble those reported in the main text.

\begin{figure}
    \centering
    \textbf{Noisy Probabilities in De Finetti's Thought Experiment:}\par\medskip
    \includegraphics[scale=1]{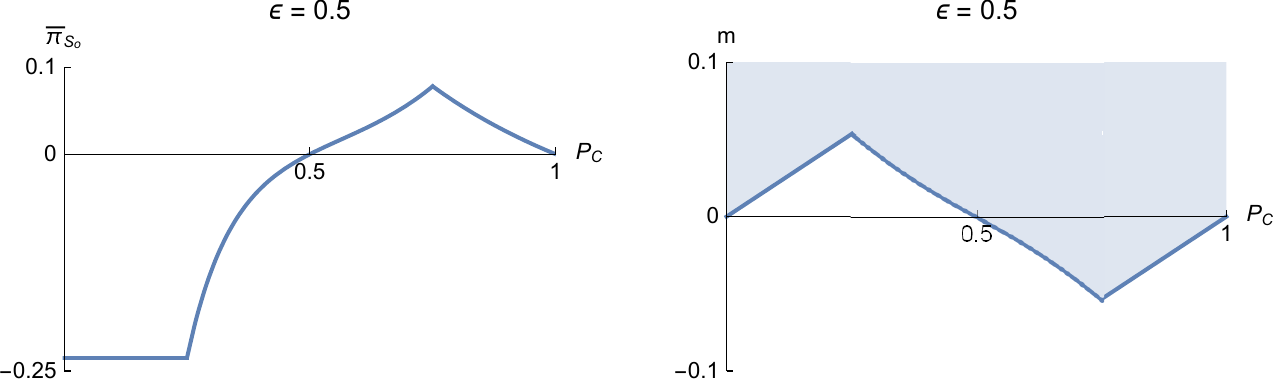}
    \caption{\normalfont The subject's mean expected margin in the role of seller in De Finetti's classic thought experiment when $m = 0$. Left: The adjustments to $P_C$ needed to establish fair odds. The shown pattern is clearly comparable to the one reported in Figure VII of the main text.}
\end{figure}


\pagebreak

\bibliography{C:/Users/uwn/Dropbox/LaTeX/References/arXiv_FOfNP}{}

\noindent \textsc{}\newpage 

\end{document}